\documentclass{ar-1col}
\usepackage[numbers,sort&compress]{natbib}
\usepackage{url}
\usepackage[version=3]{mhchem}
\usepackage{gensymb,amssymb}
\let\micro\micro
\let\perthousand\perthousand

\begin{document}

\markboth{Greenaway et al.}{Ternary Nitrides}

\title{Ternary Nitride Materials: Fundamentals and Emerging Device Applications}

\author{Ann L. Greenaway,$^1$ Celeste L. Melamed,$^{2,1}$ M. Brooks Tellekamp,$^1$ Rachel Woods-Robinson,$^{3,4,1}$ Eric S. Toberer,$^2$ James R. Neilson$^5$ and Adele C. Tamboli$^1$*

\affil{$^1$Materials and Chemistry Science and Technology Directorate, National Renewable Energy Laboratory, Golden, Colorado, United States, 80401 \\*email: adele.tamboli@nrel.gov}
\affil{$^2$Department of Physics, Colorado School of Mines, Golden, Colorado, United States, 80401}
\affil{$^3$Applied Science and Technology Graduate Group, University of California at Berkeley, Berkeley, California, United States, 97402}
\affil{$^4$Energy Technologies Area, Lawrence Berkeley National Laboratory, Berkeley, California, United States, 94720}
\affil{$^5$Department of Chemistry, Colorado State University, Fort Collins, Colorado, United States, 80523}}

\begin{abstract}
Interest in inorganic ternary nitride materials has grown rapidly over the past few decades, as their diversity of chemistries and structures make them appealing for a variety of applications. Due to synthetic challenges posed by the stability of \ce{N2}, the number of predicted nitride compounds dwarfs those that have been synthesized, offering a breadth of opportunity for exploration. This review summarizes the fundamental properties and structural chemistry of ternary nitrides, leveraging metastability and the impact of nitrogen chemical potential. A discussion of prevalent defects, both detrimental and beneficial, is followed by a survey of synthesis techniques and their interplay with metastability. Throughout the review, we highlight applications (such as solid-state lighting, electrochemical energy storage, and electronic devices) in which ternary nitrides show particular promise.

\end{abstract}

\maketitle

\tableofcontents

\section{INTRODUCTION}\label{intro}

Nitride materials surged in importance over the 20th century as numerous applications were identified from light emitting diodes, lasers, and power electronics, to superconductors and hard coatings. As the utility of nitrides increased, so did the need for broader structure-property tunability, spurring the development of ternary nitride research. The nitrogen atom in the solid state presents both a promise and a challenge: its strong bonds lead to compelling materials properties and long experimental lifetimes, but make synthesis challenging due to the relative inertness of \ce{N2}. Despite the prevalence of nitrogen on Earth \cite{zhang_distribution_1993}, there are few reported ternary nitrides compared to  ternary oxides, making this a promising field for research~\cite{zakutayev_design_2016, sun_map_2019}. 

\begin{figure}
    \centering
    \includegraphics[width=150mm]{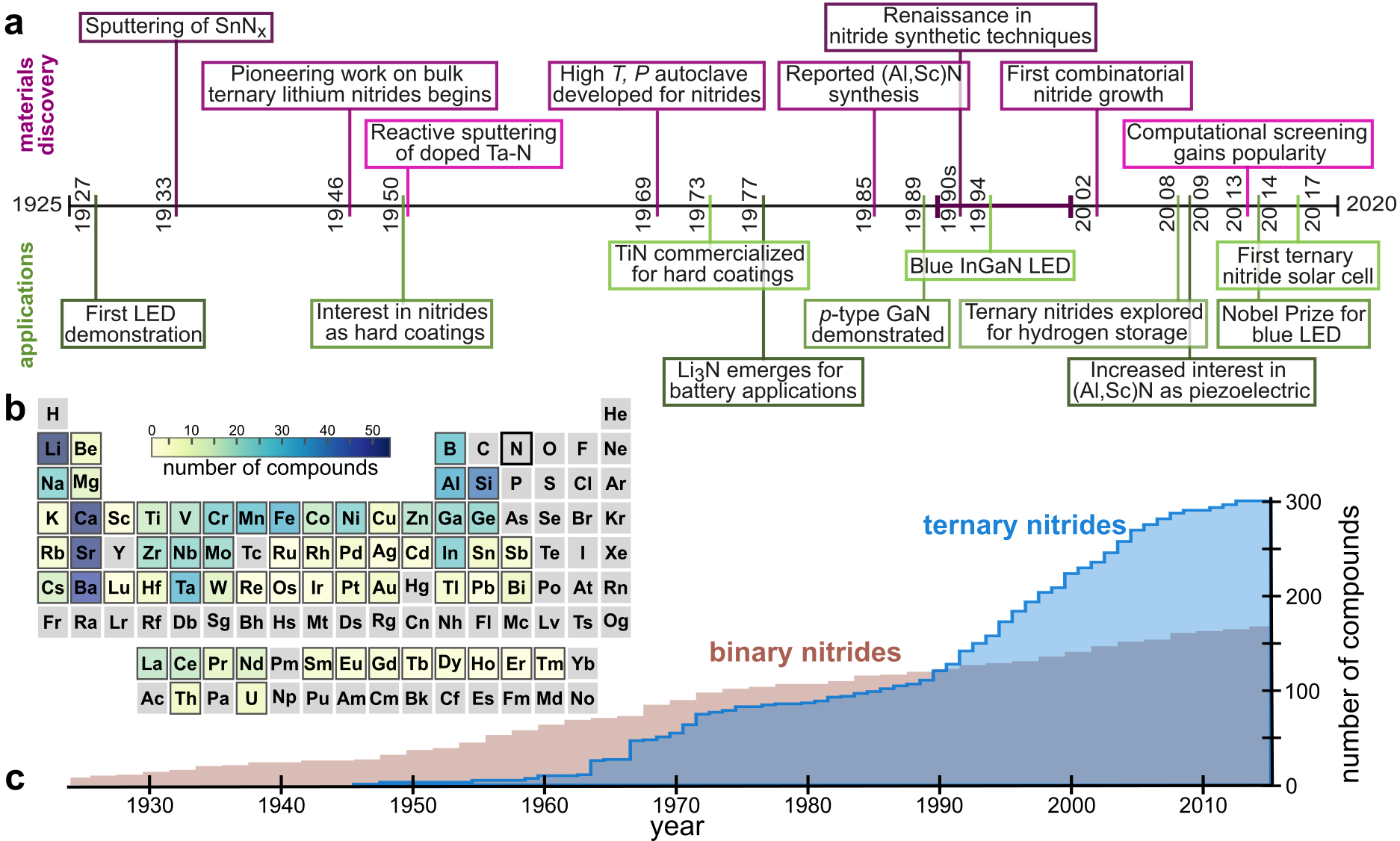}
    \caption{(a) A timeline of important events in nitride research divided between materials discovery and developments in applications. (b) A heat map of the frequency of elements in fully-ordered ternary nitrides as reported in the Inorganic Crystal Structure Database (ICSD). Compounds where nitrogen was not the sole anion were removed from the list. (c) A histogram of all experimentally-synthesized binary and ternary nitride compounds (with nitrogen as the sole anion) over time, from the ICSD. Each compound refers to a unique crystal structure, and the compound's year refers to the earliest report of each compound. Additional information for all sections can be found in Section S2 of the SI. [**Note to Annual Reviews: We created this figure for this article; it is not based on any previously published image.**]}
    \label{fig:nitrides_over_time}
\end{figure}

Ternary nitrides were sparsely studied from the 1950s through the 1980s (see Figure \ref{fig:nitrides_over_time}a,c), with early work by a handful of research groups in studies limited to synthesis and fundamental structural characterization \cite{juza_kristallstrukturen_1946, wintenberger_groupe_1973, patterson_preparation_1966, disalvo_ternary_1996}. In the 1990s, interest in ternary nitrides increased as techniques such as solid state metathesis, ball milling in a nitrogen atmosphere, microwave-generated nitrogen plasmas and novel molecular and oxide precursors were used in exploratory syntheses \cite{brese_crystal_1992, fitzmaurice_low-temperature_1993,wiley_rapid_1992,miki_preparation_1992,houmes_microwave_1997,elder_lithium_1992,schnick_nitridosilicates_1997}.\,\begin{marginnote}
\entry{Ternary inorganic nitride}{A compound consisting of two metal cations and the nitrogen anion. This review focuses on emerging \textit{compounds} in this class rather than heavily investigated \textit{alloys} such as In$_x$Ga$_{1-x}$N.}
\end{marginnote} This expansion has continued into the 21st century, with progress in both the development of previously synthesized nitrides for applications varying from hydrogen storage to photovoltaic absorbers and light-emitting diodes (LEDs) \cite{langmi_ternary_2008, javaid_thin_2018,tellekamp_heteroepitaxial_2020}, as well as the continued search for new ternary nitrides, driven by the rise of computational materials prediction and discovery \cite{hinuma_discovery_2016,sun_map_2019,zakutayev_design_2016}. Interest in ternary nitrides has led to a commensurate increase in their elemental diversity (Figure \ref{fig:nitrides_over_time}b), expanding this space to an exciting array of potential chemistries.

In this review, we highlight how the diversity of the nitride chemistry enables a wide breadth of structures and properties, reaching beyond established application spaces where binary nitrides have historically been used. New applications are emerging in which ternary nitrides can uniquely serve, such as batteries and phosphors. Many of these new potential applications are still theoretical or early stage: there is a huge opportunity space for device-oriented materials research in ternary nitrides. We highlight the unusual metastability of ternary nitrides and their ability to be integrated into hybrid devices with binary nitrides. We discuss how the particular defect chemistry of ternary nitrides plays into properties, such as cation site disorder and anion vacancies, which leads to some challenges and some opportunities for applications. 

\section{FUNDAMENTAL PROPERTIES AND STRUCTURAL CHEMISTRY}\label{chemistry}

Inorganic ternary nitrides span a broad range of compositions (Figure \ref{fig:nitrides_over_time}b) and stoichiometries, yielding a commensurate breadth of properties. This chemical diversity can be traced back to the unusual properties of nitrogen. In the following, we discuss the solid-state chemistry of nitrides, addressing the behavior of nitrogen as well as cation chemistries and metastability, and highlighting recent examples of compounds that continue to expand our understanding of these materials.

\subsection{Nitrogen in the Solid State}\label{nitrogenChem}
\begin{figure}
    \centering
    \includegraphics[width=150mm]{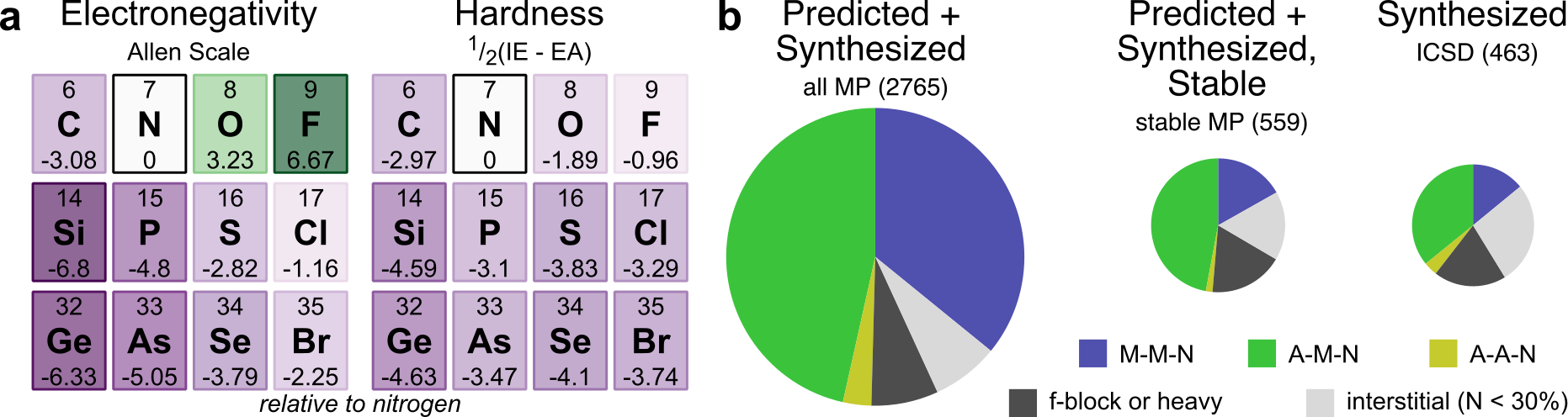}
    \caption{(a) Comparison of electronegativity and hardness of common anions relative to the values for nitrogen. (b) Comparison of chemical distributions of ternary nitrides across materials databases. Only unique compounds are counted across all of the Materials Project (MP), thermodynamically ``stable'' compounds in MP ($E_{hull} = $ 0 eV/atom) and the experimental ICSD. Further details can be found in Section S3 of the SI. [**Note to Annual Reviews: We created this figure for this article; it is not based on any previously published image.**]}
    \label{fig:nitride_classes}
\end{figure}

To understand the nitrogen anion, it is illustrative to compare it to other non-metal elements. Figure \ref{fig:nitride_classes}a highlights electronegativity and absolute hardness across the p-block elements. Nitrogen is very electronegative (comparable to Cl, but less than O and F) \cite{allen_electronegativity_1989}, with its electronegativity arising from its poorly shielded nucleus. This also drives nitrogen to be the hardest element of the anion-forming main group elements  \cite{parr_absolute_1983}. This high chemical hardness and intermediate electronegativity give rise to nitrogen’s unique properties, where it can form high-coordination ionic compounds but also form covalent bonds as a result of its large, delocalized electron cloud in the formal anion, N$^{3-}$.

\begin{marginnote} 
\entry{Chemical hardness}{Quantification of the soft/hard acid/base concept, where soft acids are acceptor atoms with large, diffuse, polarizable electron clouds and easily excited outer electrons; hard acids have the inverse properties \cite{parr_absolute_1983}.}
\end{marginnote}

This unique combination of hardness and electronegativity allows nitrogen to adopt various bond hybridizations (and therefore bonding geometries) in the solid state \cite{king_chemical_1995}, enabling coordination numbers from 2 to 8, although 6-coordinate is the most common \cite{niewa_recent_1998, disalvo_ternary_1996, brese_crystal_1992}. Nitrogen also represents a middle ground of bonding between anions: its variable bond character enables on the one hand better orbital overlap with metals than O, resulting in smaller band gap semiconductors \cite{zakutayev_design_2016}, and on the other hand compounds with  metallic conductivity similar to carbides \cite{haglund_theory_1993}. Furthermore, there are examples of nitrogen adopting multiple bond geometries within the same compound \cite{schnick_nitridosilicates_1997, marchand_nitrides_1991}. This variability in bond hybridization and mixed ionic-covalent character, as well as high polarizability of the nitrogen anion \cite{kniep_ternary_1997}, affords highly cohesive materials with a large structural diversity \cite{gregory_structural_1999, sun_thermodynamic_2017}. 

\subsection{Cation Chemistries}\label{cationChem}
Ternary nitrides form, or are predicted to form, with metals from across the periodic table, as shown in Figure \ref{fig:nitrides_over_time}b.  While elemental nitrogen has a high chemical hardness, the addition of electron density creates a rather polarizable anion, thus leading to immense structural diversity with different cations.  However, because of the challenges of ternary nitride synthesis, discussed in detail in Section \ref{synthesis}, the number of experimentally known materials in this category is dwarfed by the number of predicted compounds. Illustrating the compositional variability of ternary nitrides, Figure \ref{fig:nitride_classes}b plots the prevalence of five nitride classes across two materials databases (additional comparisons can be found in Section S3 of the SI). The first three classes are the focus of this review: “M-M-N”, “A-M-N,” and “A-A-N” follow Sun, et al. \cite{sun_map_2019}, where “A” is an alkali or alkaline earth metal and “M” is a transition or p-block metal. The statistics presented in the figure include multiple instances of compounds within the same chemistry system (e.g., \ce{Zn3MoN4} and \ce{ZnMoN2} \cite{arca_redox-mediated_2018} in the system Zn-Mo-N). The categorization “f-block or heavy” designates compounds containing any lanthanide (La--Yb) or element with atomic number greater than 83 (Bi), while the “interstitial” class designates materials with less than or equal to 20\% atomic N in their formula.

While the distribution of predicted ternary nitrides varies across databases, it is clear that A-M-N compounds form a plurality of predictions and in the experimental ICSD. Many of the earliest discovered ternary nitrides were based around strongly electropositive metals such as Li, including \ce{LiZnN} and \ce{Li3AlN2} \cite{juza_kristallstrukturen_1946}, which stabilize compounds through the inductive effect \cite{ niewa_recent_1998,sun_map_2019}. In contrast, A-A-N compounds are the smallest portion of compounds across databases, because of difficulty stabilizing multiple electropositive metal cations \cite{sun_map_2019}. The proportion of M-M-N compounds varies widely across databases, and for the MP is substantially reduced when only stable compounds are considered. However, this value is close to the number of M-M-N compounds in the ICSD, an indication of historical challenges of synthesizing ternary nitrides with less electropositive metal cations. This distribution both sheds light on historical challenges in nitride discovery and indicates rich opportunities for synthesis of new nitrides, particularly in the M-M-N space which, as will be described in Section \ref{metastability}, is generally metastable.

\subsection{Structural Chemistries}\label{TNstructuralChem}

\begin{figure}
    \centering
    \includegraphics[width=150mm]{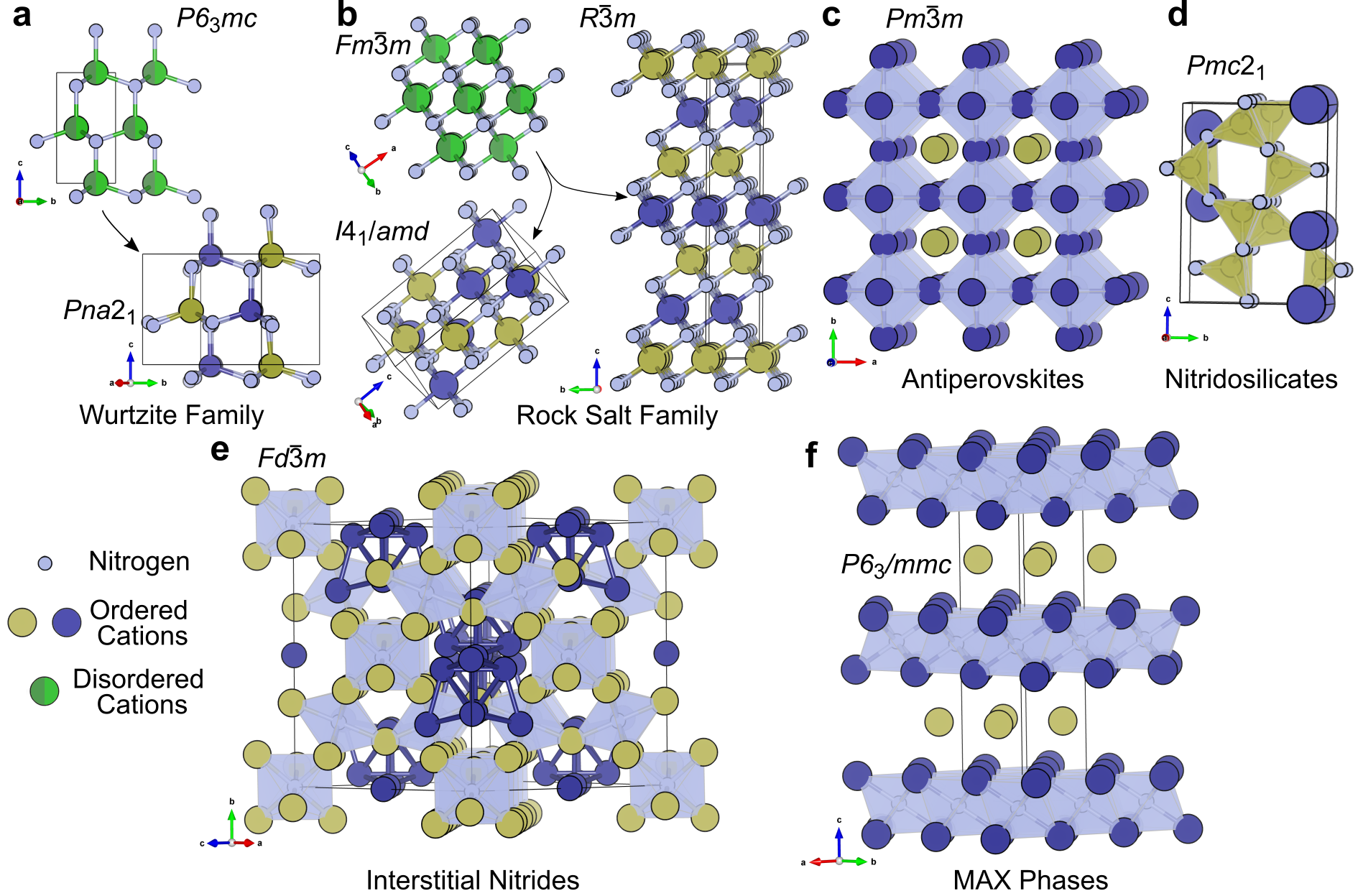}
    \caption{Some structural families of ternary nitrides. Many functional nitrides crystalize in (a) polar, tetrahedral wurtzite-derived structures with disordered ($P6_3mc$) or ordered ($Pna2_1$) configurations or (b) non-polar octahedral rock salt-derived structures with disordered ($Fm\bar{3}m$) and ordered ($I4_1amd$ or $R\bar{3}m$) configurations. Antiperovskites (c) in the $Pm\bar{3}m$ structure are also common. Nitridosilicates crystallize in myriad structures, many of which are analogous to the large family of silicate oxides; depicted here in (d) is \ce{Sr2Si5N8} ($Pmc2_1$) owing to its relevance for solid-state lighting phosphors. Also included are example structures of (e) interstitial nitrides and (f) MAX/MXene layered structures. All structures are drawn to the same scale.}
    \label{fig:structures}
\end{figure}

The combination of nitrogen’s chemical properties and the range of cations with which it can form bonds leads to wide structural variety in ternary nitrides. These factors taken together enable nitrogen to form in more diverse structural types, particularly for ternary materials, than are observed for other anions, including unique structures which do not have analogous structures in other anion systems (e.g. layered $P6_3/mcm$ \ce{Ca6$M$N5}, where $M$ = Ga, Mn, Fe \cite{gregory_structural_1999}). Figure \ref{fig:structures} shows a sampling of these structures, many of which are of particular interest for technological applications. The wurtzite family (Figure \ref{fig:structures}a) is derived from III-N materials by cation mutation \cite{pamplin_systematic_1964,xia_chemistry-inspired_2017}, and similarly the rocksalt family (Figure \ref{fig:structures}b) is analogous to binary rocksalts such as \ce{TiN}. Cation-ordered structures for these materials appear in computational structure predictions, but cation-disordered structures are often observed experimentally \cite{martinez_synthesis_2017}, as illustrated. Both of these classes are interesting for technological applications due to their potential for heteroepitaxial integration with widely used binary nitrides while enabling wider property spaces, as discussed in Sec.~\ref{integration}. Antiperovskites (Figure \ref{fig:structures}c) are common structures for ternary nitrides \cite{heinselman_thin_2019}, but constitute metallic interstitial compounds; as of this writing there are only a few computationally predicted perovskite nitrides \cite{sarmiento-perez_prediction_2015}, one of which was recently synthesized \ce{LaWN3} \cite{talley_synthesis_2020}. Ternary nitrides can also be viewed from the perspective of shared motifs, such as corner-sharing \ce{SiN4} tetrahedra which make up nitridosilicates (Figure \ref{fig:structures}d), which have been considered for applications including phosphors for LEDs (see Application Spotlight: Phosphors for Solid-State Lighting)\cite{schnick_nitridosilicates_1997}. Although not discussed in depth here, interstitial ternary nitrides such as the structure in Figure~\ref{fig:structures}e demonstrate metallic behavior as a result of being extremely metal-rich, and, among other properties, are of interest for ferromagnetic applications \cite{szymanski_prediction_2019}. Finally, MAX phases and MXenes (Figure~\ref{fig:structures}f) are increasingly common in the literature due to interest in their use for catalysis and for battery applications, as described in Application Spotlight: Electrochemical Energy Storage. Ternary nitrides can form in structures beyond those highlighted here, including a range of structures with anion-centered polyhedra (\textit{anti} structures, see Ref~\citenum{gregory_structural_1999}).

\begin{textbox}[h]\section{Application Spotlight: Phosphors for Solid-State Lighting}\label{lighting}
Ternary nitrides provide many practical advantages for phosphor and solid-state lighting applications, as recently reviewed \cite{george_phosphors_2013}.  To enable luminescence, one typically substitutes in \textit{ca.}1-5\% of a lanthanide ion with a partially-filled $4f$ electronic subshell that permits atom-like $4f$ to $5d$ absorption and re-emission transitions (e.g., \ce{Eu^{2+}}). The $5s$ and $5p$ orbitals shield the optically accessible $4f$ sub-band transitions from lattice coupling, allowing efficient optical absorption and recombination \cite{mitchell_perspective_2018}. Ternary materials provide an advantage: a stiff main group-nitride based framework  (as found in the nitridosilicates, Figure~\ref{fig:structures}d) prevents non-radiative relaxation via phonon quenching, since the relevant phonons have a low population at operation temperatures.  The nitrogen-based ligands to the lanthanide provide a larger crystal field splitting with the $5d$ energy levels than with oxides (e.g., more covalency); thus,  the nitride ligand provides more ability to tune color, often into the needed red colors.  While quantum efficiencies of top performing materials have a little room for improvement, the color rendering index and synthesis remain major challenges.  
\end{textbox}

\subsection{Metastability}\label{metastability}

In addition to their wide range of structures, nitrides have a higher propensity than other material classes to crystallize in thermodynamically metastable phases, which are crystal structures that are not the thermodynamic ground state (see sidebar). To determine $E_{hull}$, a compound's free energy referenced to its ground state energy, many studies use density functional theory (DFT) to calculate and compare phase formation energies \cite{affleck_quantum-statistical_1981}. In DFT databases and high-throughput screenings, $E_{hull}$ is typically estimated by calculating formation energy at 0 K (approximately equal to free \textit{enthalpy} $\Delta H_f$) rather than temperature-dependant free energy $\Delta G$ due to computational simplicity, though this neglects entropic effects. In this subsection, reported $E_{hull}$ values  refer to $\Delta H_f$ calculated by the DFT GGA or GGA+U formalism unless specified otherwise. It should also be noted that for a given compound to be metastable and synthesizable it must also be \textit{dynamically} stable (i.e., not have imaginary phonon modes).

\begin{marginnote} 
\entry{Energy above the convex hull, $E_{hull}$}{An energetic metric of thermodynamic stability in which, for a given solid compound, a “convex hull” is constructed by connecting the thermodynamic ground state free energies within its composition phase space \cite{anelli_generalized_2018}. $E_{hull}$ = 0 eV/atom indicates a ground state thermodynamically stable compound, while $E_{hull} >$ 0 eV/atom indicates a metastable compound. Larger values of $E_{hull}$ indicate larger degrees of metastability.}
\end{marginnote} 

\begin{figure}
    \centering
    \includegraphics[width=150mm]{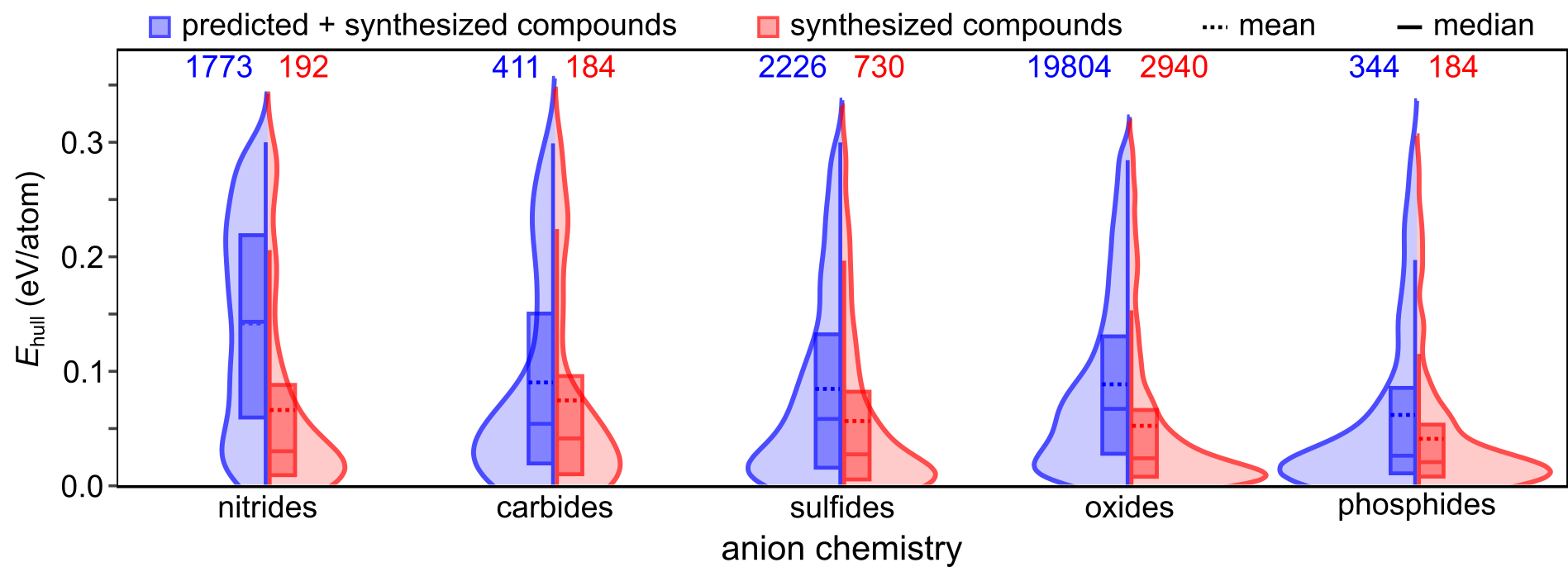}
    \caption{Violin plots comparing metastability across all metastable nitride, carbide, sulfide, oxide, and phosphide compounds in the MP database \cite{jain_commentary_2013} (see SI for details). Blue distributions represent all metastable compounds in each class and red distributions represent the  experimentally synthesized subset. Box plots describe statistics within each set, with cutoffs corresponding to lower and upper quartiles (25\% and 75\%), solid lines corresponding to medians, and dashed lines corresponding to means. Numbers above each distribution denote the number of compounds in each class. Further details can be found in Section S4 of the SI. [**Note to Annual Reviews: We created this figure for this article; it is not based on any previously published image.**]}
    \label{fig:metastability_violin}
\end{figure}

To examine metastability and synthesizability, Figure \ref{fig:metastability_violin} compares $E_{hull}$ values of predicted and synthesized metastable materials across single-anion compounds in the Materials Project database (see Supporting Information for figure details). Although materials with the lowest $E_{hull}$ values tend to be the easiest to synthesize, not all low energy polymorphs are synthesizable \cite{stevanovic_sampling_2016}. Notably, nitrides have a much larger window of metastable synthesizability than oxides \cite{sun_thermodynamic_2016}, as can be seen in Figure \ref{fig:metastability_violin} by their large range of $E_{hull}$ values.  Metastable oxides, phosphides, and sulfides have quartile $E_{hull}$ values in the range 0.01 eV/atom to 0.13 eV/atom for all compounds, while the quartile metastability of both predicted and experimental nitrides extends far higher in energy, from 0.06 to 0.22 eV/atom. Carbides have a lower predicted, but experimentally similar, metastability range than nitrides (the relationship between these two classes varies with temperature, see Ref.~\citenum{bartel_physical_2018}). It is also important to note that databases are subject to sampling bias, discussed further in Section S1 of the SI.

Various explanations for thermodynamically metastable phase stability exist in literature \cite{kroll_pathways_2003}. As discussed in Section~\ref{nitrogenChem}, nitrides form mixed ionic-covalent bonds and therefore compounds with higher cohesive energies, enabling experimental access to higher values of $E_{hull}$ \cite{sun_thermodynamic_2017}. Nitrides (and carbides) also have a significantly higher amorphous thermodynamic upper limits, the energy levels above which compounds cannot be synthesized \cite{aykol_thermodynamic_2018}. It is important to consider a compound's metastability with respect to both \textit{constituent binaries} and \textit{competing ternary polymorphs}. A ternary nitride can be metastable with respect to its binary nitride end points, but still be the lowest energy polymorph at its composition. Metastable polymorphs can also phase separate and co-exist, recently shown as competing rock salt and wurtzite derived \ce{MgSnN2} phases \cite{greenaway_combinatorial_2020}. Additional metrics for assessing (meta)stability are important to consider, such as ``energy above the Pourbaix hull'' which accounts for moisture degradation \cite{singh_electrochemical_2017, woods-robinson_assessing_2018}.

Metastability is intimately connected to experimental control of the nitrogen chemical potential $\mu$(N).  Synthesis approaches applying a  high $\mu$(N) can stabilize metastable phases that would otherwise be out-competed by binary phase segregation \cite{caskey_thin_2014,sun_thermodynamic_2017} and enable kinetic quenching of metastable phases; stabilization can also be achieved through defect engineering, alloying, and epitaxial templating \cite{siol_accessing_2019}. A primary challenge in nitride synthesis is therefore to maintain (or increase) a high $\mu$(N) through control of experimental parameters. Additionally, the methods used to control $\mu$(N) can greatly impact the defect landscape of a synthesized material. Practical control of $\mu$(N) for a variety of synthesis conditions is discussed further in Section~\ref{synthesis}.

\begin{textbox}[h]\section{Application Spotlight: Electrochemical Energy Storage}\label{energyStorage}
The structure and chemistry of ternary nitrides may offer several advantages to advance current battery technologies. As noted in Figure \ref{fig:nitrides_over_time}a, the binary $\alpha$-\ce{Li3N}, $P6/mmm$, was an early candidate material for battery applications due to its high \ce{Li+} conductivity \cite{alpen_ionic_1977} arising from 1--2 Li vacancies per unit cell which is critical to intraplanar hopping. While \ce{Li3N} is not suitable as an electrolyte due its low decomposition potential ($\sim$0.5\,V) \cite{gregory_lithium_2008}, substituting various metals for Li in order to form a ternary (\ce{Li_{3-x}M_xN}, M = Cu, Si, Co, Mn, or Fe) enables its use as an electrode material \cite{gregory_lithium_2008, liu_lithium_2004}. In fact, up to one Li vacancy per formula unit in the 8-coordinate nitride structure has been demonstrated \cite{cabana_towards_2008}.

These  promising characteristics have made ternary nitrides attractive for high-rate, high-capacity electrochemical storage as both electrolytes and negative electrodes \cite{zhu_strategies_2017}. Layered compounds are particularly of interest for ease of Li$^+$ (or other ion, see Ref. \citenum{verrelli_viability_2017, verrelli_study_2019}) intercalation, and layered \ce{LiMoN2} and \ce{Li7MnN4} \cite{balogun_recent_2015} have been investigated as electrode materials, while anti-fluorite \cite{yamane_ternary_2001} and hexagonal lithium nitridosilicates \cite{ischenko_formation_2002} have been explored as promising solid state electrolytes with measured Li-ion conductivity as high as $\mathrm{5\times 10^{-2}\Omega^{-1} cm^{-1}}$ at elevated temperatures. As seen in Figure \ref{fig:structures} and discussed in Section \ref{disorder}, some ternary nitrides can be synthesized with disorder on the cation site, which may lead to additional benefits when used as electrode materials. Disturbances to the potential landscape from cation disorder can prevent local trapping during ion transport, leading to higher ion conductivity and longer cycle stability \cite{hanghofer_substitutional_2019, saha_influence_2019}. 

Ternary nitrides have also been suggested as candidate anode coatings, contingent upon low electrical conductivity, high ionic conductivity, and tolerance for cycle-induced strain \cite{zhu_strategies_2017}. However, a primary challenge of nitride electrochemistry is material stability, partially due to the tendency of nitride electrodes to undergo irreversible conversion reactions to form more stable phases rather than reversible intercalation at room temperature \cite{balogun_recent_2015}. Ternary nitrides MAX and MXene phases may also see development for other electrochemical device applications, including supercapacitors. Although MXenes are very new materials \cite{anasori_2D_2017}, they are the focus of intense research, particularly the nitride compounds which are underrepresented compared to MXene carbides.  
\end{textbox}

\section{DEFECTS}\label{defects}

One of the most challenging yet potentially revolutionary aspects of ternary nitrides is their propensity for complex defects. In this section, we discuss the intrinsic, extrinsic and extended defects commonly found in ternary nitrides, and the challenges and opportunities posed therein.

\subsection{Intrinsic Defects}\label{intrinsic}
    
Intrinsic defects are historically pernicious in nitride materials. Though self-interstitials and antisite defects tend to be high energy in binary nitrides, both cation and anion vacancies often cause undesirable compensation of dopants. Cation vacancies become detrimental for wide-band-gap binary nitrides in particular because defect formation energies typically decrease for increasing band gap \cite{van_de_walle_defects_1999}. Many of these issues carry over to the ternary nitrides, and are made more complex by the sheer number of intrinsic defects enabled by a ternary system. 

\begin{marginnote} 
\entry{Defect formation energy, $E^f[X^q]$}{The energetic cost to create or remove an isolated defect $X$ with charge state $q$ from a bulk material. Defect formation energy calculations are used to assess how favorable various defects and dopants are to form compared to one another.}
\end{marginnote} 

\begin{figure}
    \centering
    \includegraphics[width=150mm]{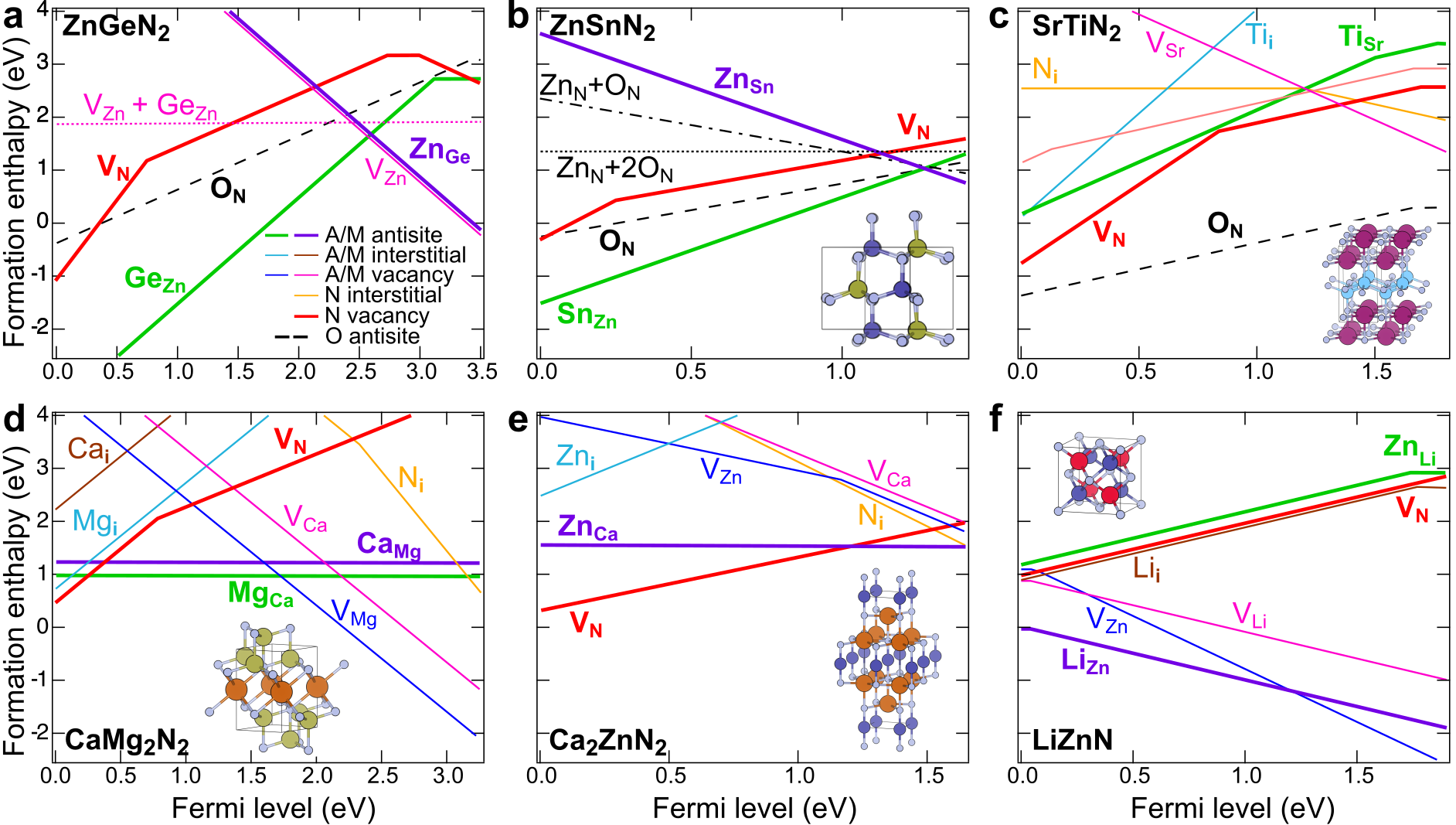}
    \caption{A sampling of defect diagrams, where the Fermi level axis refers to the band gap, for different inorganic ternary nitrides: (a) \ce{ZnGeN2} (b) \ce{ZnSnN2} (c) \ce{SrTiN2} (d) \ce{CaMg2N2} (e) \ce{Ca2ZnN2} (f) \ce{LiZnN}. Structures are inset for all materials except \ce{ZnGeN2}, which is isostructural to \ce{ZnSnN2}. Further details can be found in Section S5 of the SI. Data in panel (a) from Reference~\citenum{melamed_combinatorial_2020}. Data in panel (b) from Reference~\citenum{pan_interplay_2019}.  Data in panel (c) from Reference~\citenum{he_intrinsic_2019}. Data in panels (d-e) from Reference~\citenum{hinuma_discovery_2016}. Data in panel (f) from Reference~\citenum{toyoura_first-principles_2007}. [**Note to Annual Reviews: We created this figure for this article; it uses data from previously published papers but the figures are unique.**]}
    \label{fig:defectDiagrams}
\end{figure}

As computational work on ternary nitrides has increased, defect diagrams have emerged as a way to predict and examine the effects of defects in emerging materials, particularly target compounds for optoelectronic applications. Defect formation energy diagrams visualize the formation energy or enthalpy ($E^f$, see sidebar) of intrinsic and extrinsic defects as a function of Fermi energy, enabling prediction of defect formation, equilibrium Fermi level and dopability. A defect's slope indicates its charge, with positive slope indicating an electron donor and negative slope indicating an electron acceptor, and defects with lower $E^f$ are more likely to form. $E^f$ diagrams for six ternary nitrides spanning a range of structure types are shown in Figure \ref{fig:defectDiagrams}. Figure~\ref{fig:defectDiagrams}a-b shows two M-M-N compounds, which are wide band gap (ZnGeN$_2$) and a narrower band gap (ZnSnN$_2$) examples of the wurtzite-derived II-IV-N$_2$ family. SrTiN$_2$, shown in (c), is an A-M-N ternary nitride with the same stoichiometry but a layered structure. The materials shown in (d-e) have cations of the same valence but different stoichiometries than (a-c): the A-A-N compound \ce{CaMg2N2} and A-M-N compound \ce{Ca2ZnN2}. Finally, LiZnN (f) is an A-M-N in Zn and N form a zincblende network and Li occupies tetrahedral interstitials. Diagrams were selected to show chemical potential regimes representative of common synthetic conditions: for example, all diagrams were calculated in N-rich regimes, and cation chemical potentials were chosen to reflect vapor pressure conditions during thin-film, vacuum synthesis, e.g. Zn-poor. These diagrams represent a variety of bonding configurations, chemistries, and structure but clearly cannot capture defect trends for all ternary nitrides. 

Nitrogen vacancies (V\textsubscript{N}) are consistently low $E^f$ in ternary nitrides just as in binaries. As shown in Figure~\ref{fig:defectDiagrams}, V\textsubscript{N} is the lowest $E^f$ intrinsic donor for \ce{SrTiN2}, \ce{CaMg2N2}, and \ce{Ca2ZnN2} (c-e). V\textsubscript{N} in \ce{ZnGeN2} and \ce{ZnSnN2} have similarly low $E^f$ (ranging from -1\,eV at the valence band maximum to 3\,eV near the Fermi level for \ce{ZnGeN2}), but are not the lowest $E^f$ intrinsic donor due to more favorable cation antisite defects (discussed further below). Similarly, cation vacancies tend to be low $E^f$ in the selected ternary nitrides. In particular, V\textsubscript{Zn} is the lowest $E^f$ intrinsic acceptor in \ce{ZnGeN2} and is low $E^f$ in \ce{Ca2ZnN2} and \ce{LiZnN}, and V\textsubscript{Mg} is low $E^f$ in \ce{CaMg2N2}. Though low $E^f$ vacancies may negatively impact properties and/or cause undesired defect compensation of extrinsic dopants, they are not always detrimental: for example, Li vacancies are necessary for ionic charge transport in battery applications, see Applications Spotlight: Electrochemical Energy Storage \cite{gregory_lithium_2008}.

Native antisite defects play a defining role in the defect landscape of ternary nitrides. Antisite defects tend to be higher $E^f$ for ternary nitrides with dissimilar cation atomic sizes such as SrTiN$_2$, in which neither antisite defect has a $E^f$ below 0\,eV (Figure ~\ref{fig:defectDiagrams}c). However, for ternary nitrides where the two cations are close in size, such as wurtzite-derived \ce{ZnGeN2} and \ce{ZnSnN2} (Figure~\ref{fig:defectDiagrams}a-b), cation antisite defects such as Zn\textsubscript{Sn} and Zn\textsubscript{Ge} have low $E^f$. 
The effect of donor doping is to push the Fermi level higher in the band gap, necessarily decreasing the formation energy of compensating electron acceptors, and the opposite is true of acceptor doping \cite{yan_doping_2008}. Thus, compensation presents a challenge for applications where extrinsically doped ternary nitrides are desirable. For example, native cation antisite defects are predicted to compensate ZnGeN$_2$ when doping both n-type~\cite{adamski_optimizing_2019} and p-type ~\cite{adamski_strategies_2019}. However, it can also be a benefit, e.g. in ZnSnN$_2$ where self-compensation of electron donors with Zn\textsubscript{Sn} defects enables n-type doping control~\cite{fioretti_combinatorial_2015}. Similarly, in LiZnN (Figure~\ref{fig:defectDiagrams}f), the Li\textsubscript{Zn} antisite is the lowest $E^f$ defect at energies close to the valence band maximum and indicates this material is intrinsically p-type~\cite{toyoura_first-principles_2007}. There are experimental strategies to prevent defect compensation, including changing $\mu$(N)~\cite{tsunoda_electrically_2018} and co-doping with hydrogen or other extrinsic dopants (discussed in Section~\ref{extrinsic}). Selection of materials with significant cation size mismatch would result in a larger energetic driving force toward cation order \cite{martinez_synthesis_2017}, such as in (Zn,Mg)SiN$_2$ compounds compared to (Zn,Mg)GeN$_2$. When antisites are favorable, intentional use of cation off-stoichiometry may offer a route to purely intrinsic bipolar doping. Additionally, in some ternary nitrides cation antisite defects can form complexes enabling structural tunability, namely the use of cation site disorder as a knob to tune properties. 

\subsection{Cation disorder} \label{disorder}
Cation antisite defect pairs, as discussed above, can be prevalent in ternary nitrides. Antisite defect complexes can exist well beyond the dilute defect approximation typically used in calculations, and at high concentrations are referred to as \textit{cation site disorder}. Cation-disordered ternary nitrides can be thought of as occurring in higher-symmetry structures with a generic cation site that has partial occupancy for each cation, as shown in Figure~\ref{fig:structures}a,b. This type of disorder typically \cite{nakatsuka_orderdisorder_2017, schnepf_using_2020} narrows the band gap (but not always \cite{bauers_ternary_2019}). It is not yet clear what the nature of this modification is: whether a mid-gap state is introduced that may become band-like at high concentration \cite{haseman_deep_2020}, near-band edge ``tail" states which effectively narrow the band gap \cite{lany_monte_2017}, or some other mechanism that impacts the electronic structure. While its origins are not fully understood, modifying the band gap through cation disorder is a promising mechanism for band gap tuning in optoelectronic devices, and cation disorder may impact other properties such as thermal conductivity and ion conductivity as recently reviewed in Ref.~\citenum{schnepf_utilizing_2020}. Challenges to disorder-based tunability remain, include precisely controlling the amount of cation site disorder, accurately characterizing and quantifying disorder, and creating theoretical predictions for realistic, metastable structures.

\begin{textbox}[h]\section{Application Spotlight: Optoelectronic Devices}\label{optoelectronics}
One of the most promising proposed applications for ternary nitride semiconductors is in the active regions of light emitting devices such as LEDs and lasers, or conversely in the absorber layer of solar cells. There is a need for a wider portfolio of optoelectronic semiconductors that can be integrated with known semiconductors (see Sec.~\ref{integration}), and ternary nitrides offer a suite of possibilities. Proposed LED device architectures utilize the predicted type II band offsets between ZnGeN$_2$ and GaN to achieve longer wavelength light emission in the ``green gap" region~ \cite{hyot_design_2019, han_designs_2016}. For solar cells, ZnSnN$_2$ has been studied as a thin film photovoltaic material comprised of Earth-abundant, inexpensive elements~\cite{fioretti_combinatorial_2015}, and an initial device demonstration has been reported~\cite{javaid_thin_2018}.

Cation disorder is an integral component to understanding optoelectronic applications of ternary nitrides, both through defect considerations and enhanced tunability. Because cation antisite defects are so significant in ternary nitrides, their impact on properties must be considered when designing materials for devices such as LEDs~\cite{haseman_deep_2020} and photovoltaics~\cite{lany_monte_2017}, where recombination and minority carrier lifetimes are critical factors. Defects can lead to non-radiative recombination pathways through trap-assisted recombination, but defect localization and energy of the trap state within the band are important in determining radiative/non-radiative recombination lifetimes. Localized defect states result in lattice distortions, increasing electron-lattice coupling strength and thus increasing the probability of phonon-mediated recombination and decreasing the probability of radiative recombination \cite{das_what_2020}. On the other hand, due to the direct relationship between localized lattice distortions and electron-lattice coupling, defect states which are less localized and more band-like suppress trap-assisted non-radiative recombination \cite{luque_understanding_2012, luque_intermediate_2006}. 

With this in mind, understanding the electronic structure of antisite defects and cation disorder is crucial to the development of ternary nitrides for optoelectronic device applications. Defects become band-like when their concentration exceeds the Mott threshold \cite{gunning_negligible_2012}; antisite defects could potentially form as complexes with concentrations exceeding the Mott threshold in cation-disordered cases, leading to the suppression of trap-assisted non-radiative recombination. This assumes that such band-like states will form deep within the band gap, which has been suggested based on experimental mid-gap cathodoluminesence in \ce{ZnGeN2} \cite{haseman_deep_2020}. Further complicating the picture, some defect complexes are uncharged (zero slope in a defect diagram), and the energy at which they charge (change slope) may not be within the band gap, exemplified by the V\textsubscript{Zn}-Ge\textsubscript{Zn} complex in Figure~\ref{defects}a. Such energetically favorable intrinsic defects should be carefully examined when considering a ternary nitride for optoelectronic applications, including defect position within the band, defect concentration, and the effect of compensating species on defect level. 

Cation-disorder-induced band gap narrowing, if controllable, can be a powerful tool for optical device engineering, potentially enabling LEDs with disordered quantum wells and ordered barriers of the same composition, or low-cost photovoltaic devices with a disorder-tunable band gap~\cite{javaid_band_2018, schnepf_utilizing_2020}. Integration of ternary nitrides possessing tunable band gaps with mature systems like GaN, as discussed in Section~\ref{integration}, would allow layers to be picked for optimal optical function without regard to dopability, relying on the mature system for carrier injection \cite{schnepf_utilizing_2020,tellekamp_heteroepitaxial_2020,karim_effects_2020}. 
\end{textbox}

\subsection{Extrinsic Defects}\label{extrinsic}
Unintended extrinsic dopants are a major problem for both binary and ternary nitride synthesis. Unintentional oxygen contamination is a crucial issue across all nitrides due to the close atomic sizes of oxygen and nitrogen.  Oxygen contamination is difficult to avoid, even in very pure growth environments \cite{van_de_walle_defects_1999}.  In III-Ns, oxygen is a shallow donor and historically caused n-type conductivity in GaN~\cite{van_de_walle_defects_1999}. This issue persists in the ternary nitrides \cite{tsuji_heteroepitaxial_2019,melamed_blue-green_2019}, as visualized in Figure~\ref{fig:defectDiagrams}a-c. If oxygen is present, O\textsubscript{N} substitutional defects are low $E^f$, even at high $\mu$(N). Consequently, oxygen presents a huge challenge to p-type doping of ternary nitrides and developing synthesis techniques to minimize oxygen incorporation is crucial \cite{tsuji_heteroepitaxial_2019,adamski_hybrid_2017}.
However, oxygen defects can be beneficial: O\textsubscript{N} substitutional defects can form complexes with acceptors such as cation antisite defects; for example, in \ce{ZnSnN2}, Zn\textsubscript{Sn} antisites and O\textsubscript{N} defects compensate to create charge-neutral defect complexes, helping to lower the degenerate carrier concentration~\cite{fioretti_exciton_2018}; more information on \ce{ZnSnN2} can be found in Ref. \citenum{khan_review_2020}. Though oxynitride structures are outside of the scope of this review, anion site alloying with oxygen may present opportunities for new structures and properties~\cite{pan_perfect_2020, melamed_blue-green_2019}.
    
Another relevant extrinsic defect is hydrogen, which is often present due to hydrogen-containing nitrogen sources (ex. NH$_3$) and can be used to tune the defect landscape in semiconductor materials. Hydrogen (which typically occupies interstitial sites) tends to suppress the formation of compensating nitrogen vacancies in III-Ns, and then can be removed with a post-growth anneal, ``activating" p-type dopants~\cite{van_de_walle_defects_1999}. The H\textsubscript{i}-Mg complex in this case has a lower formation energy than other compensating defects, allowing compensation to occur without occupation of a lattice site. In turn, interstitial H is easily removed with a post-growth anneal, activating the Mg. This same effect has been demonstrated in ternary nitrides: growth in a hydrogen environment followed by a post-deposition anneal has been shown to passivate acceptors in \ce{ZnSnN2}~\cite{fioretti_effects_2017} and is theorized to do the same in \ce{ZnGeN2}, but H passivation is an open area of research in other ternary nitrides~\cite{adamski_strategies_2019}.
While intentional extrinsic doping of ternary nitrides will be critical for their utilization in some applications, this research is largely limited to computational work \cite{he_intrinsic_2019,adamski_optimizing_2019,adamski_strategies_2019} and experimental investigation of such dopants is currently an underexplored area of research.

\subsection{Extended Defects}\label{structuralDefects}
As discussed in Section~\ref{metastability}, metastable polymorphs are prevalent in nitrides. The zincblende polymorph of GaN lies only a few meV above the wurtzite ground state \cite{lei_heteroepitaxy_1993}, leading to prevalent inclusions of zincblende in normally wurtzite material, or stacking faults. For ternary nitrides, the presence of another cation increases the polymorph diversity, allowing for multiple cation-ordered polytypes within the same structural family and for polymorphs of entirely different structure types. Like other II-IV-\ce{N2} family compounds, \ce{MgSnN2} forms in a wurtzite-derived orthorhombic structure. Here, the orthorhombic unit cell is a 2$\times$2$\times$1 supercell of wurtzite cells with a slight bond distortion from the heterovalent cations. One cation site-swap per unit cell produces two antisite defects, but two cation site-swap actions produce a different space group---changing from $Pna2_1$ to $Pmc2_1$. An inclusion of one space group within another can be thought of as a stacking fault, but one that retains the wurtzite-derived structure rather than a cubic (zincblende) inclusion. The effect of this type of stacking fault is predicted to be electronically benign as long as the local octet rule is satisfied~\cite{quayle_charge-neutral_2015, lany_monte_2017}. The energy difference between the two wurtzite-derived space groups for \ce{MgSnN2} is only 5 meV/atom, and two similarly-related zincblende structures lie 25--30 meV/atom above the ground state, suggesting that stacking faults similar to GaN may occur in this material. However, to date only wurtzite-type and substantially more metastable rocksalt \ce{MgSnN2} (70--100 meV/atom) have been observed \cite{greenaway_combinatorial_2020,kawamura_synthesis_2020}; the coincident formation of these polymorphs likely results in similar extended defects to stacking faults.  In addition, there may be inhomogeneous regions of more or less cation-disordered phases, which could result in potential fluctuations.

\section{SYNTHESIS OF TERNARY NITRIDES}\label{synthesis}

The synthesis of ternary nitride materials revolves around the challenge of maintaining a high $\mu$(N) while also enabling mass transport and crystal growth. Here, we begin with an overview of $\mu$(N), and consequently discuss synthetic methods according to the dominant method of controlling $\mu$(N): bulk processes are classified by control via pressure, while thin film processes are classified by deposition chemistry. Despite the differences between $\mu$(N) control in bulk and thin film methods, there are cross-cutting challenges for the synthesis of ternary nitrides across these techniques. In addition to the discussion of $\mu$(N) presented here, several reviews discuss synthesis techniques of ternary nitrides in further detail \cite{tareen_mixed_2019,gregory_structural_1999}.

\subsection{Controlling Nitrogen Chemical Potential, $\mu$(N)}\label{chemPot}

\begin{marginnote}
 \entry{Chemical Potential}{Defined as the change in free energy upon addition or removal of a component while keeping pressure ($p$), temperature ($T$), and the other species ($n_{j \neq i}$) constant:\newline \newline $\mu_i = \left(\frac{\partial G}{\partial n_i}\right)_{p,T,n_{j \neq i}}$ \newline \newline which leads to its total differential:\newline  $d\mu = V dp - S dT$,  \newline where \textit{V} is volume and \textit{S} is entropy. \ce{O2}  has a lower chemical potential compared to \ce{N2} at standard conditions: \newline $\mu$(\ce{O2}) $= -0.318$ eV/atom versus $\mu$(\ce{N2})$= -0.298 $ eV/atom \cite{bale_factsage_2002}. Oxides are more stable vs \ce{O2} than nitrides are vs \ce{N2}, $\Delta H_f(0\text{K})$: \ce{Si3N4} $-1.3$ eV/atom vs \ce{SiO2} $-3.3$ eV/atom \cite{jain_commentary_2013} when referencing gases in their standard state. }
 \end{marginnote}
 
\begin{figure}
    \centering
    \includegraphics[width=150mm]{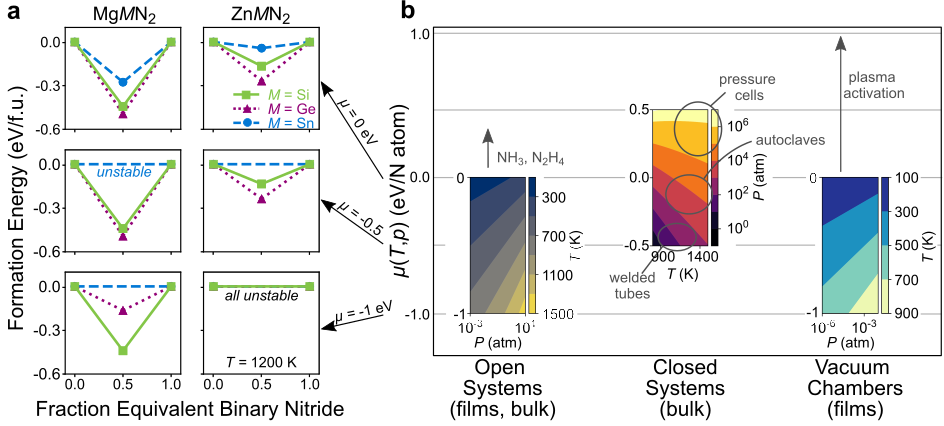}
    \caption{(a) Thermodynamic convex hull constructions using Gibbs Free Energy illustrate how shifting $\mu$(N) changes the stability of \ce{Mg\textit{M}N2} and \ce{Zn\textit{M}N2} compounds (\textit{M} = Si, Ge, and Sn). Colors correspond to $M$ species. All calculations run at $T=$ 1200~K for ($\mu $= 0 eV/atom (top), $-0.5$ eV/atom (middle), and $-1.0$ eV/atom (bottom)); values are calculated using the Gibbs predictor in a Grand Potential \cite{bartel_physical_2018,ong_python_2013}. Elevated $\mu$(N) is required to stabilize \ce{Zn\textit{M}N2} compounds compared to \ce{Mg\textit{M}N2} for a given \textit{M} cation. (b) Approximate mapping of synthesis methods onto the chemical potential of nitrogen with respect to standard conditions, $\mu(T,p) = \mu(T,p_0) + kT\ln(p/p_0)$, where $p_0 = 10^5$ Pa or 1 bar and assuming ideal gas behavior: open systems, closed systems at elevated pressure, and closed systems at low pressure (vacuum systems). Use of \ce{NH3} or \ce{N2H4} can result in a higher effective chemical potential if one assumes that atomic nitrogen is produced rather than \ce{N2} \cite{sun_thermodynamic_2017}. In the synthesis of thin films (or in some bulk reactions \cite{vajenine_plasma-assisted_2007}), a plasma discharge can increase the effective chemical potential by nearly $\sim$1 eV/atom \cite{caskey_semiconducting_2015}.  
    Across these synthetic regimes, $\mu(N)$ is tunable by more than 2 eV/atom. Further details found in Section S6 of the SI. [**Note to Annual Reviews: We created this figure for this article; it is not based on any previously published image.**]}
    \label{fig:metasynth}
\end{figure}

Figure~\ref{fig:metasynth}a highlights how the synthesis of ternary nitrides is sensitive to $\mu$(N), as shown for the Mg- and Zn- based ternary nitrides. The conditions used to determine stability in Figure~\ref{fig:metasynth}a are explicitly connected to common synthetic conditions in Figure~\ref{fig:metasynth}b. With open systems, increasing temperature decreases $\mu$(N); as such, one may employ  lower temperatures using more aggressive chemical reactants (e.g., \ce{NH3}, \ce{N2H4})  in order to retain a high effective chemical potential while promoting chemical reactivity. While atmospheres of \ce{NH3} and \ce{N2H4} have enhanced reactivity relative to \ce{N2}, they do not have an elevated $\mu$(N) by rigorous definition due to their equilibrium decomposition at elevated temperature to \ce{N2} and \ce{H2} \cite{allison_janaf_1996}. Regardless, these species are more kinetically active (e.g., due to their lack of triple-bonded N) and can yield thermodynamically-spontaneous reactions due to the formation of \ce{H2} or \ce{H2O} as byproducts. In contrast to open systems, closed systems generate elevated $\mu$(N) through the use of pressure. Vacuum chambers permit additional modification of $\mu$(N) based on the ability to generate various nitrogen-based plasmas. Plasmas can contain a mixture of atomic nitrogen species or metastable excited molecules with an elevated $\mu$(N) \cite{merel_influence_1998}, and their composition can have a significant influence on growth rate \cite{iliopoulos_active_2005}, surface morphology, and defect concentration \cite{clinton_observation_2019}. 

\subsection{Bulk Synthesis and Crystal Growth}\label{bulkGrowth}

As shown in Figure~\ref{fig:nitrides_over_time}a, the earliest demonstrations of ternary nitride materials relied on bulk synthesis \cite{juza_kristallstrukturen_1946}, and these techniques have continued to play a major role in the development of this class. Bulk syntheses enable materials discovery with the ability to control over crystallization and cation ordering which may not be available in thin-film geometries. Bulk processes are useful for synthesizing powder and single crystalline material which can be used to investigate their crystal structures and basic optical, thermodynamic, and magnetic properties. We classify bulk syntheses as being in an open system, which has a practically infinite reservoir of nitrogen at a constant $\mu$(N), or as being in a closed system that has a finite amount of nitrogen such that $\mu$(N) may evolve over the course of a reaction.

\subsubsection{Open Systems, Ambient Pressure}

Synthesis with open chemical systems often relies on the gas-phase nitridation of precursor materials with a range of nitrogen sources. For electropositive s-block, lanthanide, or early transition metals, \ce{N2} is sufficiently oxidizing to support nitride formation, as in the synthesis of \ce{\textit{Ln}Si3N5} (\textit{Ln} = lanthanide) from biphasic lanthanide-silicon alloys \cite{schnick_nitridosilicates_1997} or \ce{MgMoN2} from the reaction of \ce{N2} with \ce{Mg3N2} and Mo \cite{verrelli_viability_2017}. Others include \ce{LiMgN} \cite{yamane_ternary_2001}, \ce{Li2Zr2N2}, \ce{Li2HfN2} \cite{niewa_re-evaluation_1995}, \ce{Ca2TiN4}, \ce{Ca5NbN5}  \cite{hunting_synthesis_2007}, \ce{Li7MnN4}\cite{cabana_ex_2005}, \ce{MgTa2N3}, \ce{Mg2Ta2N4}, \ce{CaTaN2} \cite{verrelli_study_2019}, and \ce{ScTaN2} \cite{niewa_metal-metal_2004}. 
Other solid-state reactions, such as direct reaction of the binary nitrides, can proceed under \ce{N2} atmosphere at high temperature if the formed compound is sufficiently stable as to overcome the reduced $\mu$(N) at elevated temperature, such as the production of \ce{MgSiN2} from its constituent binaries \cite{bruls_preparation_1999}. Some reactions are even robust enough to proceed in ambient atmospheres with the right precursors, such as the flame-propagated synthesis of \ce{MgSiN2} from the metathesis of \ce{Mg3N2} and \ce{SiO2} \cite{blair_solid-state_2005}, although with  oxygen inclusion in the product. 

Ammonolysis reactions based on traditional solid-state chemistry are convenient, although they require temperatures $>$ 900 \degree C, for the atomic-scale mixing of metals within a ternary oxide host, as in the reaction \ce{Zn2GeO4 + 3 NH3  -> ZnGeN2 + Zn(g) + 4 H2O(g) + 1/2 N2(g) + 1/2 H2(g)} \cite{maunaye_preparation_1970} or for production of \ce{LiMoN2} from \ce{Li2MoO4} \cite{elder_thermodynamics_1993}. \ce{NH3} serves as a reactive, activated N source, but in contrast to \ce{N2} is fairly reducing from the equilibrium decomposition of \ce{NH3} into \ce{N2} and \ce{H2}.  Creative gas plumbing has been used to delay the ammonia decomposition at high temperature by keeping the gas cool before contacting the solid powders \cite{otsuka_synthesis_2016}. While the reaction of oxides in ammonolysis yields an additional thermodynamic benefit from producing \ce{H2O(g)} as a side product, the incorporation of oxygen in the final product is a substantial concern (see Section~\ref{extrinsic}); however, ammonolysis is not limited to oxides as source material \cite{tessier_original_1997}. 

Reactive N-containing precursors in open systems backed  with near-ambient partial pressures of \ce{N2} and \ce{NH3} maintain a sufficient $\mu$(N) to drive synthesis of nitrides. Silicon diimide (\ce{Si(NH)2}), an industrial precursor for \ce{Si3N4}, can be reacted directly with alkaline earth or lanthanide metals at temperatures in excess of 1500 \degree C to form very stable phases such as \ce{CeSi3N5}\cite{lange_silicon_1991,schnick_nitridosilicates_1997}. Likewise, metal amides serve to increase reactivity, as found in the production of \ce{Na3WN3} from \ce{W2N} with excess \ce{NaNH2} \cite{rauch_synthesis_1994}. However, in all cases with open systems, it remains difficult to react the more electronegative metals (e.g., Sn) or avoid the inclusion of oxygen-based impurities without special purification procedures. 

\subsubsection{Closed Systems, Elevated pressure}

Closed systems are especially appropriate for reactions of binary nitrides, where all nitrogen needed for the final product is embedded in the precursor materials. These reactions are well-suited for ternaries where one or more metal is an alkali or alkaline earth metal due to the availability of N-containing precursors (e.g., \ce{Li3N}, \ce{Mg3N2}, \ce{Sr2N}, and \ce{Ca3N2} \cite{verrelli_viability_2017,hunting_synthesis_2007,yamane_ternary_2001}), and can be extended to \ce{NaN3}-based compounds with appropriate safety precautions for azide use (see sidebar). Reactions of binary nitrides are typically contained in welded, refractory metal tubes to avoid reaction of the metals with other containers (e.g., as occurs with \ce{SiO2)} and to retain a reasonable $\mu$(N) for the endogenously-released N. However, for lower-temperature reactions ($T <$ 1000 K), ion exchange or double exchange reactions can take place in an inert-atmosphere glovebox \cite{gillan_rapid_1994, gillan_synthesis_1996} or an evacuated \ce{SiO2} ampoule \cite{zakutayev_experimental_2014}. 

\begin{marginnote}[]\entry{Safety}{All synthetic routes described here present their own hazards, but many are \textit{more} hazardous than might be expected given the benign nature of nitrogen itself. Inorganic azides, \ce{N3^-}, are acutely toxic and extremely hazardous, especially as they are prone to decomposition by release of \ce{N2}. Supercritical \ce{NH3} is accessible above 405 K/132 \degree C and 11.3 MPa/110 bar, requiring special considerations for such high $T$ and $P$. Carcinogenic \ce{N2H4}, which can be a N source for bulk and thin-film syntheses, decomposes spontaneously in the presence of many metal catalysts, including some of the metals identified in Figure \ref{fig:nitrides_over_time}b.}
\end{marginnote}

Welded tubes are also used for reactions performed in molten, metallic fluxes, which have garnered success for crystal growth and materials discovery efforts. The molten metal serves not only to enhance mass transport, but also to increase $\mu$(N) by dissociation of \ce{N2} into the metal, creating an activated species. These methods have been employed for GaN synthesis, where use of Na flux rather than Ga self-flux reduces the needed pressure from $\sim10^4$ atm of pressure at 1500 \degree C \cite{karpinski_equilibrium_1984} to 50 atm of \ce{N2} at 750 \degree C \cite{aoki_growth_2001}. Specific to the synthesis of ternary nitrides with alkaline earth metals, it has been shown that inclusion of Ca, Sr, and Ba in molten sodium can increase the solubility of nitrogen to $\sim$1.1 mol\% N attributed to the formation of species such as ``\ce{[Ba4N]}'' \cite{addison_reaction_1975,yamane_sodium_2018}. The use of alkali or alkaline earth metal fluxes \cite{yamane_sodium_2018} (including eutectic mixtures) has led to the growth of \ce{Ba2ZnN2} and \ce{Sr2ZnN2} \cite{yamane_synthesis_1995}; \ce{Ba3Ga2N4}, \ce{Ba5Si2N6} and \ce{Ba3Ge2N2} \cite{yamane_barium_1996}; and \ce{Ca6Te3N2} \cite{dickman_metal_2016}, as well as higher-order multinary nitrides.

\subsubsection{Closed Systems, Extreme Pressures}

A powerful approach in the discovery and crystal growth of ternary nitrides has been through ammonothermal reactions held in specially designed autoclaves. Using corrosion and high-$T$-resistant alloys (e.g., nickel-based superalloys), reactions can be conducted in supercritical \ce{NH3} at high temperatures ( $>$ 600 \degree C) and pressures ($>$ 600 MPa or 6 kbar) \cite{wang_ammonothermal_2006, richter_chemistry_2014,hausler_ammonothermal_2018}. The high-mobility, high $\mu$(N) supercritical fluid provides an ideal environment for synthesis, and the solution chemistry can be carefully tuned using various mineralizers (e.g., acidic, \ce{NH4$X$}; basic, \ce{$A$NH2}). This tuning is vital for the successful formation of ternary nitrides \cite{richter_chemistry_2014} and can be used to enable or suppress intermediate species or defects (e.g., mixed-metal amides and hydride formation, respectively). The supercritical ammonia approach has been used with the basic mineralizer \ce{KNH2} to synthesize of \ce{ZnSiN2} and \ce{ZnGeN2} with strong evidence of cation ordering at $T$= 870-1070 K, $P \le$ 230~MPa or 2.3 kbar  \cite{<hausler_ammonothermal_2018a>}. These methods are ideally suited for fundamental studies that require high-quality single crystals, as well as exploratory synthesis, and are reviewed in depth elsewhere \cite{hausler_ammonothermal_2018}.  However, the expense and expertise required to mitigate hazards of these processes have prevented widespread adoption.

Multi-anvil presses, while also specialized and expensive, feature fewer hazards. $\mu$(N) is high above GPa-scale hydrostatic pressures at $T>$1000 K ($\mu–\mu(T_0,p_0) > 0$ eV/atom, Figure~\ref{fig:metasynth}b), and extreme $T$ and $P$ also increase the (normally prohibitively slow) solid-state diffusion \cite{borg_introduction_1988} needed for the crystallization and cation ordering of many ternary nitrides \cite{endo_high-pressure_1992,hinuma_discovery_2016}. For instance, while the metathesis reaction between \ce{ZnCl2}, \ce{Si3N4} and \ce{LiN3} in a welded Ta tube at 700 \degree C yields nano-sized grains of $Pna2_1$ \ce{ZnSiN2} with cation ordering, annealing at 6 GPa and 1200 \degree C is required for full crystallization of the material \cite{endo_high-pressure_1992}. In some cases, a judicious choice of precursor is required to ensure atomic-scale mixing of the reaction, as in the metathesis of fluoride salt precursors to crystallize \ce{ZnSnN2} at 5~GPa \cite{kawamura_synthesis_2016}.  The low throughput and significant expense of extreme pressure reactions has reserved their use for specialized, basic science studies.

\subsection{Thin Film Synthesis}\label{filmSynthesis}

The requirement for thin-film geometries in many device applications has driven the widespread use of physical and chemical vapor deposition (PVD and CVD) techniques and an increasing role for these techniques in materials discovery. Thin film synthetic methods have simplified the measurement of certain properties, such as electronic transport and optical phenomena, which are difficult to directly probe using bulk material and are crucial for many device applications. While we divide techniques between common PVD and CVD types, there are many thin film growth methods which do not fall neatly in to either category \cite{du_synthesis_2008,quayle_synthesis_2013,quayle_vapor-liquid-solid_2017,grekov_methods_2004}. Additional disucssion of these techniques can be found in Ref.~\citenum{tareen_mixed_2019}.

\subsubsection{Physical Vapor Deposition}
PVD methods including reactive sputtering and plasma-assisted molecular beam epitaxy (MBE) have been extremely useful for the discovery of ternary nitrides. The success of these techniques derives mainly from the use of plasma N sources, which adds another variable besides  $T$ and $p$ to  $\mu$(N) (Figure~\ref{fig:metasynth}b); it should be noted that modelling and predicting the resulting $\mu$ depends on various and complex factors pertaining to the plasma (see SI).  
The plasma composition strongly influences the properties of the deposited material, both in phase selection and defect formation. Plasma activation may be necessary to stabilize some materials (see Fig.~\ref{fig:metasynth}). However, high ion content plasma (e.g., electron cyclotron resonance) can increase defect density in films through impact damage when ions are accelerated by an induced floating potential \cite{clinton_observation_2019} and increased ion concentration in the plasma can lead to higher defect concentrations and poor optical properties \cite{ptak_relation_1999, blant_nitrogen_2000}. Materials with lower bond strength (comparatively lower band gaps) like \ce{ZnSnN2} are more susceptible to plasma damage. In sputtering, accelerated ions in the plasma are necessary to knock source atoms from the targets. Accelerating potentials in sputtering are often much higher than in MBE, leading to ions impacting the growing surface with up to 40 eV of kinetic energy \cite{thornton_influence_1974}. Various methods can be used to decrease ion damage in sputtering such as increasing operating pressure (increasing gas-phase collisions) or using decelerating substrate potentials, pulsed high-frequency signals \cite{arakawa_high_2016}, and off-axis geometries \cite{umeda_improvement_2006}. In other cases, ion bombardment can provide beneficial energy for crystallization and enhancement of gas reactivity \cite{howson_reactive_1994}.

 A major benefit of PVD methods is the relative simplicity of the techniques and precursors, which enables rapid mapping of thin-film deposition space through combinatorial synthesis. Rapid screening of deposition conditions (e.g., metal flux, $T$) onto a stationary substrate with intentionally nonuniform temperature effectively enables many deposition experiments to occur at the same time. While this approach is most often used in reactive sputtering, it has also been demonstrated for other PVD and some CVD techniques. Although combinatorial approaches are sometimes considered to produce lower-quality material (due to high deposition rates and varying composition), some recent works have demonstrated heteroepitaxy \cite{melamed_blue-green_2019,greenaway_combinatorial_2020}, suggesting that this approach can be used to identify growth conditions appropriate for high-quality thin-film growth as well as materials characteristics.

The ease of obtaining activated \ce{N2} via plasma in PVD methods and the lack of competing side reactions given has facilitated demonstrations of new ternary nitrides. However, avoiding plasmas in PVD is also possible, for example using ammonia-MBE; the trade-off is that high substrate temperatures are required to crack \ce{NH3}. This can conflict with the need to deposit many ternary nitrides at low temperature, driven by the desorption of high vapor pressure elemental sources (e.g., Zn and alkali or alkaline earth metals) \cite{lide_crc_1993,greenaway_combinatorial_2020}. Zn-containing compounds such as \ce{ZnGeN2}, \ce{ZnSnN2}, and \ce{CaZn2N2} must be deposited below 500 \degree C in order to prevent Zn desorption~\cite{tellekamp_heteroepitaxial_2020, feldberg_growth_2013, fioretti_combinatorial_2015, melamed_combinatorial_2020,tsuji_heteroepitaxial_2019}. Higher temperatures would be beneficial to improve film morphology and crystallinity, and to reduce defect concentrations. 

\subsubsection{Chemical Vapor Deposition}
In contrast to PVD, CVD methods rely on the high activity and reactivity of gas-phase precursors and gas-phase thermodynamic byproducts (e.g. hydrocarbons, \ce{H2(g)} and \ce{HCl(g)}) to yield nitride product formation. A major benefit of this approach is the larger pressure and temperature windows for deposition afforded by the ability to choose appropriate precursors and supply them in the gas phase. Here, $\mu$(N) is indirectly controlled by the reactivity of the N-containing source, which may be \ce{N2} plasma or \ce{NH3}, or less commonly \ce{N2} or  \ce{N2H4}\cite{mizuta_low_1986,koukitu_thermodynamic_1999}. In metal-organic chemical vapor deposition (MOCVD, also called metal-organic vapor phase epitaxy, MOVPE), additional control is available in the form of various metal-organic sources. For instance, \ce{ZnGeN2} has been grown by MOCVD using diethyl Zn with \ce{NH3} and \ce{GeH4} \cite{zhu_epitaxial_1999,karim_effects_2020}, as well as using remote N plasma and \ce{Ge(CH)H3} \cite{misaki_epitaxial_2004}. Although MOCVD and the related process hydride vapor phase epitaxy (HVPE) are prominent growth methods for thin films of III-N binaries and alloys, there are very few reports of other ternary nitrides synthesized by these methods, with the above examples and a lone report of HVPE of polycrystalline \ce{ZnGeN2} from metallic sources and \ce{N2} \cite{larson_synthesis_1974}. If suitable precursors can be identified for metals in a given ternary nitride, these techniques can likely be adapted to afford the high-quality growth of thin films of such materials. For exploratory synthesis, however, precursor cost and safety hazards present a substantial barrier to entry, particularly for MOCVD.

\subsection{Integration with Binary Compounds}\label{integration}
\begin{figure}
    \centering
    \includegraphics[width=150mm]{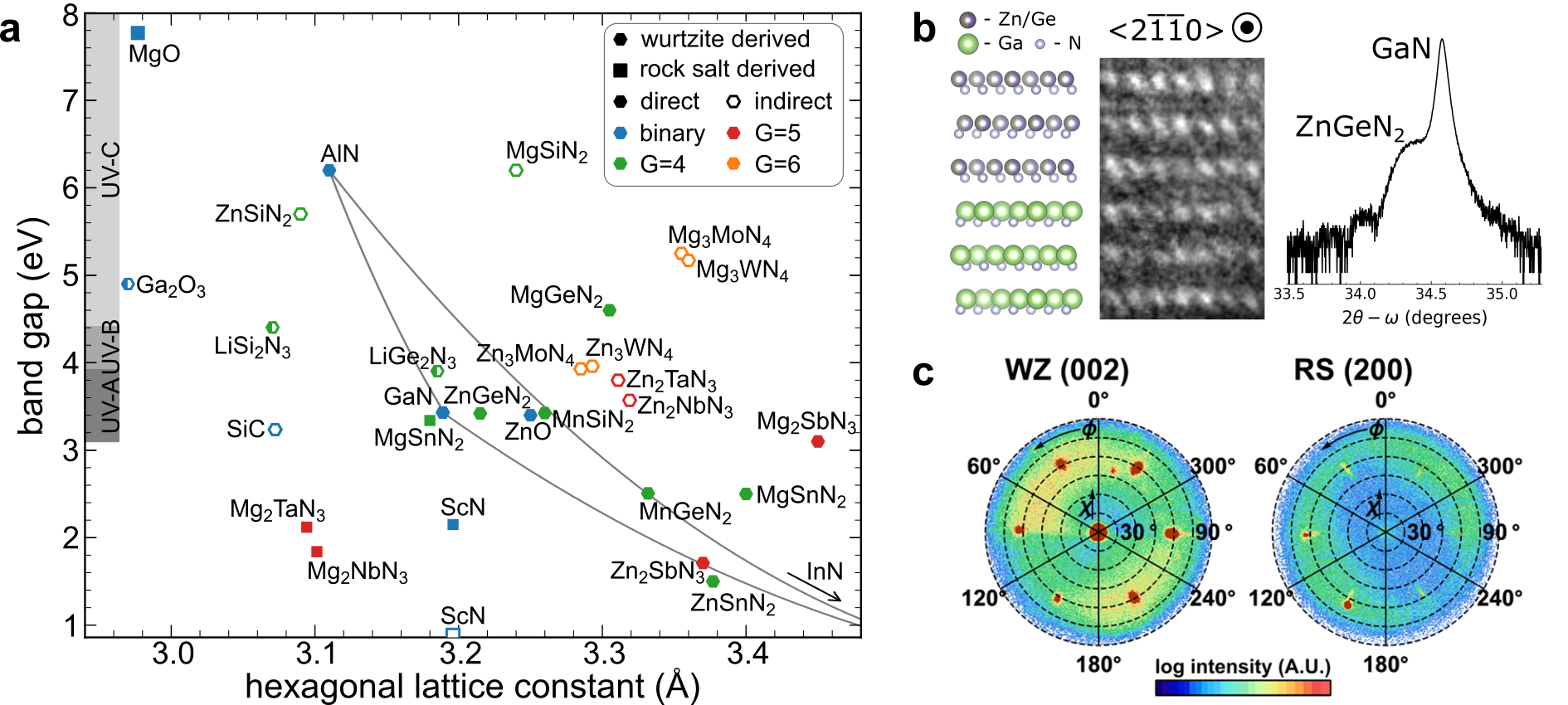}
    \caption{(a) Band gaps and lattice constants of selected wide band gap rock salt and wurtzite-derived ternary nitrides, where G is the oxidation state of the second cation, along with selected binary compounds from the same range. Demonstrations of binary nitride/ternary nitride heteroepitaxial integration have been experimentally demonstrated in a few compounds, including \ce{ZnGeN2} (b) and \ce{MgSnN2} (c). Half-shaded data points represent materials with similar computed direct and indirect transitions. Further details found in Section S7 of the SI. (b) shows a crystallographic model and corresponding high resolution scanning transmission electron microscopy of a \ce{ZnGeN2}/GaN heterointerface, along with XRD demonstrating a smooth surface and interface through coherent thickness fringes~\cite{tellekamp_heteroepitaxial_2020}. (c) shows XRD pole figures of \ce{MgSnN2} sputtered on GaN. The pole figures are taken along the wurtzite-derived (002) scattering plane and the rock salt derived (200) scattering plane, showing the stabilization of both phases~\cite{greenaway_combinatorial_2020}. Panel (b) reprinted with permission from Reference~\citenum{tellekamp_heteroepitaxial_2020}. Copyright 2020 American Chemical Society. Panel (c) reprinted with permission from Reference~\citenum{greenaway_combinatorial_2020}. Copyright 2020 American Chemical Society. [**Note to Annual Reviews: ALG and MBT are authors of these articles; the publisher grants authors the right to reuse their own figures.**]}
    \label{fig:ZnGeN2_MBE}
\end{figure}

As discussed throughout this review, ternary nitrides often have structures derived from binaries which can enable heteroepitaxial integration. For example, wurtzite-derived ternaries can be used as active layers in devices to expand the application space of binary III-N compounds. Opportunities include the use of ternary nitride emitter layers in LEDs (see Application Spotlight: Optoelectronics spotlight), hybrid high electron mobilty transistor (HEMT) devices (see Application Spotlight: Ultra-Wide Band Gap Electronic Devices), contact layers, photovoltaic absorber layers, lower-cost substrates, cladding layers with different refractive indices, and more. Most heteroepitaxy work has focused to date on known systems with Zn or Mg cations: \ce{ZnSnN2} \cite{feldberg_growth_2013},  \ce{ZnGeN2}, and \ce{MgSnN2} have each been integrated with GaN, as shown in Figure \ref{fig:ZnGeN2_MBE}b and c \cite{tellekamp_heteroepitaxial_2020, karim_effects_2020, greenaway_combinatorial_2020}. However, there is a huge opportunity space for development of new materials.

The starting point for exploring materials for integration with binaries is to consider the M$_{x}$-M$_{y}^{G+}$-N$_{z}$ space, where $G$ is the oxidation state of the second metal M. These materials tend to be either wurtzite-derived or rocksalt-derived structures. We map the band gaps and hexagonal lattice constants of some promising ternary nitride compounds along with relevant binary compounds in  Figure~\ref{fig:ZnGeN2_MBE}. DFT formation enthalphy calculations using the $E_{hull}$ formalism were used to generate thermodynamic ground states and determine if rocksalt- or wurtzite-derived compounds are shown, and GW calculated band gaps are reported maintaining the GGA(+U) wavefunction \cite{lany_band-structure_2013}. Heteroepitaxial integration can potentially occur along a vertical line with finite width (within approximately 1\% lattice constant). Not shown are many promising materials that have been predicted but not yet experimentally reported: for example, \ce{MnSnN2}, \ce{LiVN2}, and \ce{CaSnN2} are in Materials Project and may form in wurtzite-derived structures compatible with III-Ns.

Synthetic challenges in realizing these heterostructures include symmetry differences of ordered and disordered ternaries, volatility, and reactivity. When nucleating a lower-symmetry structure (ordered ternary) on a higher-symmetry one (binary or disordered ternary), there are two translationally equivalent possible nucleations, leading to threading dislocations at grain boundaries \cite{sands_stable_1990}. Nucleating high-symmetry onto low-symmetry generally does not lead to additional extended defects. In addition, many of the cations comprising ternary nitrides are highly volatile, which may limit growth temperatures and ability to overgrow ternary layers with binaries. Finally, some of the cations may be reactive with oxygen, placing constraints on growth methods.

\begin{textbox}[h]\section{Application Spotlight: Ultra-Wide Band Gap Electronic Devices}\label{electronics}
Wide band gap semiconductors form the basis of emerging power electronic devices, where the width of the band gap informs the critical field for dielectric breakdown (E$_C$) that drives figures of merit for high power devices \cite{baliga_semiconductors_1982, coltrin_transport_2017}. Nitride semiconductors are of great interest here because of their high mobilities, ultra-wide band gaps (UWBG, $>$~3.4~eV), and polarization, enabling HEMTs. As shown in Figure \ref{fig:ZnGeN2_MBE}a, many UWBG ternary nitrides exist which may be integrable with known materials and substrates. It is possible that new ternary nitrides can be integrated into III-N electronic devices or even support entire devices, but those have yet to be demonstrated.

The polarization properties of wurtzite-based ternary nitrides are especially interesting in conjunction with traditional binary nitrides or alloys. The recently proposed \ce{ZnSiN2}/\ce{AlN} polarization-induced HEMT has a nearly-lattice-matched heteroepitaxial interface with a large polarization discontinuity \cite{adamski_band_2020} that induces an electric field, accumulating carriers into a nominally undoped channel with extremely high mobility due to the absence of scattering centers. Such a device could exceed the performance of current \ce{AlGaN}/\ce{GaN} HEMTs where the lattice mismatch between AlN and GaN limits the thickness and Al-content of the polarization gate. Some UWBG ternary nitride device architectures may be limited by their dopability: ternary nitrides with low-enthalpy antisite defects will exacerbate issues related to self-compensation of extrinsic dopants in UWBG materials, as discussed in Sec.~\ref{defects}. However, as discussed in Sec.~\ref{integration}, device architectures can be envisioned that do not rely on doping control in ternary layers.

Another ternary nitride of note for electronic applications is AlScN (which can be considered an alloy, but is included here because of the dissimilar crystal structures of its binary end-members). Wurtzite AlScN has found success as a piezoelectric material in the last decade \cite{akiyama_enhancement_2009}, and recently demonstrated ferroelectric switching \cite{fichtner_alscn_2019}. Its experimental performance combined with the compatibility of AlScN with existing electronic materials indicates great promise for polarization-based applications. In addition to spontaneous polarization in the non-centrosymmetric wurtzite structure, it has been predicted that a family of distorted perovskite nitrides will show ferroelectric behavior, with polarity modifiable by an applied electric field \cite{sarmiento-perez_prediction_2015}. Recently, the ferroelectric perovskite ternary nitride \ce{LaWN3} has been experimentally demonstrated \cite{talley_synthesis_2020}. The addition of ferroelectric (and magnetic) materials to the ternary nitride material-property landscape will allow the fabrication of devices with new and useful applications \cite{jena_new_2019}.
\end{textbox}

\subsection{Challenges and Promise in Ternary Nitride Synthesis}\label{synthesisChallenges}

Because the synthesis of ternary nitrides is a relatively young field, there remain substantial challenges to its development. As depicted in Figure \ref{fig:metasynth}, established synthesis methods span a range of $\mu$(N); however, increasing $T$, which is needed for mass transport in bulk methods and to improve material morphology and crystallinity in thin-film methods, universally decreases $\mu$(N). New, innovative approaches to maintaining a high (effective) $\mu$(N) and/or support nitrogen reactivity at lower $T$ should gain significant traction going forward. As highlighted throughout this section and our discussion of defects, oxygen incorporation in particular is problematic in both bulk and thin-film ternary nitride synthesis, but many reports of new materials do not adequately control or even document the presence of oxygen. As discussed above, some bulk synthesis methods result in oxygen incorporation from the precursors; oxygen incorporation from deposition chamber components or contaminated sources is pernicious in thin-film methods, and thin films are also prone to post-growth oxidation. Together, these issues should be a major focus of development for both bulk and thin-film synthetic methods going forward. 

Where there are challenges, there is also opportunity, and we therefore highlight some developing fields for ternary nitride synthesis. The first is a trend toward bulk synthesis without control of $\mu$(N) using pressure: given the importance of $\mu$(N) in controlling nitride formation, various approaches are gaining traction in basic research, such as the use of reactive fluxes and metathesis (e.g., carbodiimides~\cite{unverfehrt_versatility_2009,hosono_melting_2019} and amides~\cite{odahara_self-combustion_2019}), as well as the use of reactive gases outside of traditional thin-film deposition chambers (e.g., \ce{N2H4}-based atomic layer deposition \cite{alvarez_enabling_2017} or \ce{N2}-based plasma \cite{houmes_plasma_1996,houmes_microwave_1997}). Another is the specific selection of substrates in order to stabilize metastable phases during deposition, an exciting possibility given the opportunities for heteroepitaxial integration discussed above. This approach has already proven successful for alloys of AlN and ScN \cite{saha_rocksalt_2018}.

\section{CONCLUSIONS}\label{conclusion}
Ternary nitrides have been steadily gaining interest as new compounds with compelling properties are predicted and synthesized. In particular, the pace of predictions has accelerated, meaning that many interesting compounds have yet to be experimentally realized. This array of possibilities stems from the diversity of nitride chemistry, including the ability of N to form bonds ranging from covalent to ionic and to support many coordination numbers, factors which in turn lead to a breadth of ternary nitride structures and properties. Many of these materials are metastable, yet synthesizable since their metastability often reflects the stability of \ce{N2} as a decomposition product. 

Defects both limit and enable device applications for ternary nitrides. As with binaries, ternaries have low energy nitrogen vacancies, which can be problematic in light of the difficulty in enhancing nitrogen reactivity in synthesis. Compounding this issue is the favorability of oxygen replacement of nitrogen, leading to pernicious unintentional oxygen incorporation. Cation antisite defects and vacancies are also favorable and can have positive or negative consequences depending on the desired properties. Cation site disorder may be useful as a method for tuning properties if it can be properly controlled and characterized, and cation vacancies can enable ion transport for applications such as batteries. However, cation site flexibility can limit doping control, as many of these materials self-compensate.

A major challenge facing ternary nitride materials discovery and application development is the difficulty in synthesizing high quality material due to the extreme stability of \ce{N2}. Many synthesis techniques have been developed to address this issue, in particular using ``active" nitrogen such as plasma sources, ammonia, hydrazine, or other reactive N-containing materials which react much more easily to form desired compounds. Temperature and pressure also provide effective knobs for tuning deposition methods, and in particular, ternary nitrides benefit from extremely high-pressure synthesis methods similar to those used for more traditional binary nitrides. Both thin film and bulk deposition techniques have been developed, including high throughput methods such as combinatorial film growth, to address the need for exploratory synthesis in this relatively new field. 

Several crucial advances are needed as ternary nitrides are further explored. New high-throughput synthetic methods should be developed which allow control of a wide range of cation precursors while maintaining high $\mu$(N) in order for experiment to keep pace with computational predictions of new materials. Highly reactive, but very pure, precursors are needed for both bulk and thin-film growth methods to reduce oxygen incorporation while facilitating synthesis of metastable ternaries. High-quality syntheses should be pursued to enable true property interrogation; to that end, methods which enable characterization of antisite defects and cation disorder across length scales and in multiple material forms (powder, single-crystal, thin-film) are needed to provide feedback to synthesis. This should ultimately enable control of intrinsic doping by cation stoichiometry and low defect density ternaries, especially if antisite defect formation is to be suppressed. Finally, the development of new substrates which can enable high-quality epitaxial growth of ternary nitrides and even provide phase stabilization will be needed in order to achieve the potential of these materials across device applications. 

Although a fairly new material class, promising applications for ternary nitrides have already emerged. Phosphors for solid state lighting have been developed based on nitrodosilicates, where the stiff framework limits phonon-related losses. The existence of Li-based ternary nitrides has spurred interest in these materials for battery applications, even beyond Li-ion.  The existence of many ternary nitrides with ultra-wide band gaps ($>$~3.4~eV) and polar crystal structures means that ternary nitrides have great potential for electronic devices such as high-frequency and high-power electronics; however, difficulties related to doping may prove challenging or insurmountable due to self-compensation. Optoelectronic applications are also fertile application spaces for ternary nitrides, and devices such as Earth-abundant photovoltaics and visible-spectrum LEDs currently motivate research in this field. For light emitting and electronics applications in particular, the ability to form heterostructures with III-N materials may ease the path to implementation of ternary nitrides by alleviating practical concerns such as doping and forming electrical contacts. While some of these applications are still in their conceptual stages, predictions have motivated increasing research towards device applications.

\section*{DISCLOSURE STATEMENT}
The authors are not aware of any affiliations, memberships, funding, or financial holdings that
might be perceived as affecting the objectivity of this review. 

\section*{ACKNOWLEDGMENTS}
Primary support for this work was provided by the U.S. Department of Energy, Office of Science, Basic Energy Sciences, Materials Sciences and Engineering Division. This work was authored in part by the National Renewable Energy Laboratory, operated by Alliance for Sustainable Energy, LLC, for the U.S. Department of Energy (DOE) under Contract No. DE-AC36-08GO28308. ALG acknowledges support from the Director’s Fellowship within NREL’s Laboratory-Directed Research and Development program. RWR acknowledges support from the UC Berkeley Chancellor's Fellowship and the National Science Foundation (NSF) Graduate Research Fellowship under Grant Numbers DGE1106400 and DGE175814. JRN thanks Matt McDermott for insightful discussions and acknowledges funding from the NSF DMR-1653863 and a Sloan Research Fellowship from the A.~P.~Sloan Foundation. EST was supported by the NSF under grant no. DMR-1555340. The authors thank Vanessa Meschke for assistance with the periodic table in Figure~\ref{fig:nitrides_over_time}. 

%


\bibliography{references} 
\bibliographystyle{ar-style3} 

\end{document}